\documentclass[runningheads]{llncs}

\newcommand*{\ShowReferences}{} 

\newcommand{\doctitle}{Relationship Estimation Metrics for Binary SoC Data}

\usepackage[pdftex,
            pdfauthor={Dave McEwan},
            pdftitle={\doctitle},
            pdfsubject={\doctitle},
            pdfkeywords={system-on-chip,
                         correlation,
                         similarity,
                         binary time series,
                         bounded time series},
            pdfproducer={},
            pdfcreator={pdflatex},
            breaklinks]{hyperref}

\def\BibTeX{{\rm B\kern-.05em{\sc i\kern-.025em b}\kern-.08em
    T\kern-.1667em\lower.7ex\hbox{E}\kern-.125emX}}

\usepackage[utf8]{inputenc}

\usepackage{pdflscape}

\usepackage{grffile} 

\ifdefined\NoGRAPHICX \else
\usepackage{graphicx}
\graphicspath{{./share/img/}}
\fi

\usepackage[plain]{fancyref}
\vrefwarning 

\usepackage{textcomp}

\usepackage{algorithmic}

\ifdefined\NoXCOLOR \else
\usepackage[usenames,svgnames]{xcolor}
\fi

\ifdefined\NoENUMITEM \else
\usepackage{enumitem}
\fi

\usepackage{placeins} 

\usepackage{multirow}
\usepackage{makecell} 

\ifdefined\NoTIKZ \else
\usepackage{tikz}
\usetikzlibrary{shapes, shapes.arrows}
\fi

\usepackage{glossaries}
\makenoidxglossaries

\newglossaryentry{naiive}{
    name=na\"{\i}ve,
    description={is a French loanword (adjective, form of naïf)
        indicating having or showing a lack of experience, understanding or
        sophistication}
}

\newglossaryentry{ImageNet}{
    name={ImageNet},
    description={is an ongoing research effort to provide researchers around
        the world an easily accessible image database.}
}

\newglossaryentry{DutyFactor}{
    name={Duty Factor},
    description={Fraction of period where signal is high.}
}

\newacronym{ad}{AD}{Adaptive Design}
\newacronym{ai}{AI}{Artificial Intelligence}
\newacronym{amba}{AMBA}{Advanced Microcontroller Bus Architecture}
\newacronym{api}{API}{Application Prognamming Interface}
\newacronym{arm}{ARM}{Acorn RISC Machine Holdings Plc}
\newacronym{ascii}{ASCII}{American Standard Code for Information Interchange}
\newacronym{aws}{AWS}{Amazon Web Services}
\newacronym{axi}{AXI}{Advanced/ARM eXtensible Interface}
\newacronym{asic}{ASIC}{Application Specific Integrated Circuit}
\newacronym{bn}{BN}{Belief Network}
\newacronym{bnn}{BNN}{Binarized Neural Network}
\newacronym{bsc}{BSC}{Binary Symmetric Channel}
\newacronym{cam}{CAM}{Content Addressable Memory}
\newacronym{cdf}{CDF}{Cummulative Density Function}
\newacronym{cisc}{CISC}{Complex Instruction Set Computer}
\newacronym{cmos}{CMOS}{Complementary Metal Oxide Semiconductor}
\newacronym{cnn}{CNN}{Convolutional Neural Network}
\newacronym{cpd}{CPD}{Conditional Probability Distribution}
\newacronym{cpu}{CPU}{Central Processing Unit}
\newacronym{dag}{DAG}{Directed Acyclic Graph}
\newacronym{ddr}{DDR}{Double Data Rate}
\newacronym{dft}{DFT}{Discrete Fourier Transform}
\newacronym{dftst}{DFT}{Design For Testability}
\newacronym{Dkl}{$D_{\text{KL}}$}{Kullback Leibler Divergence}
\newacronym{dma}{DMA}{Direct Memory Access}
\newacronym{dsp}{DSP}{Digital Signal Processing}
\newacronym{dtft}{DTFT}{Discrete Time Fourier Transform}
\newacronym{ea}{EA}{Evolutionary Algorithm}
\newacronym{eeg}{EEG}{electroencephalograph}
\newacronym{evc}{EVC}{EVent Configuration}
\newacronym{evcx}{EVCX}{EVent Configuration eXpanded}
\newacronym{evs}{EVS}{EVent Samples}
\newacronym{fec}{FEC}{Forward Error Correction}
\newacronym{ff}{FF}{Flip-Flop}
\newacronym{fft}{FFT}{Fast Fourier Transform}
\newacronym{fifo}{FIFO}{First In First Out}
\newacronym{ops}{Op/s}{Operations Per Second}
\newacronym{fir}{FIR}{Finite Impulse Response}
\newacronym{foss}{FOSS}{Free Open Source Software}
\newacronym{fpga}{FPGA}{Field Programmable Gate Array}
\newacronym{ga}{GA}{Genetic Algorithm}
\newacronym{hdl}{HDL}{Hardware Description Language}
\newacronym{hls}{HLS}{High Level Synthesis}
\newacronym{huels}{HLS}{Hue-Lightness-Saturation}
\newacronym{hmm}{HMM}{Hidden Markov Model}
\newacronym{html}{HTML}{HyperText Markup Language}
\newacronym{ic}{IC}{Integrated Circuit}
\newacronym{iff}{iff}{if and only if}
\newacronym{iid}{IID}{Independent Identically Distributed}
\newacronym{iir}{IIR}{Infinite Impulse Response}
\newacronym{ip}{IP}{Intellectual Property}
\newacronym{iqr}{IQR}{Interquartile Range}
\newacronym{isa}{ISA}{Instruction Set Architecture}
\newacronym{jtag}{JTAG}{Joint Test Action Group}
\newacronym{kde}{KDE}{Kernel Density Estimation}
\newacronym{mac}{MAC}{Media Access Control}
\newacronym{macc}{MAcc}{Multiply-Accumulate}
\newacronym{md}{MarkDown}{MarkDown Markup Language}
\newacronym{mesi}{MESI}{Modified/Exclusive/Shared/Invalid}
\newacronym{mimo}{MIMO}{Multiple In, Multiple Out}
\newacronym{ml}{ML}{Machine Learning}
\newacronym{mm}{MM}{Minimax}
\newacronym{mmu}{MMU}{Memory Management Unit}
\newacronym{lcd}{LCD}{Liquid Crystal Display}
\newacronym{lfsr}{LFSR}{Linear Feedback Shift Register}
\newacronym{lha}{LHA}{Left Hand Associative}
\newacronym{lhs}{LHS}{Left Hand Side}
\newacronym{lrm}{LRM}{Language Reference Manual}
\newacronym{lut}{LUT}{Look Up Table Cell}
\newacronym{nn}{NN}{Neural Network}
\newacronym{ocp}{OCP}{Open Core Protocol}
\newacronym{od}{OD}{Original Design}
\newacronym{ooo}{OoO}{Out-of-Order}
\newacronym{pca}{PCA}{Principle Component Analysis}
\newacronym{pcb}{PCB}{Printed Circuit Board}
\newacronym{pcie}{PCIe}{PCI Express}
\newacronym{pdf}{PDF}{Probability Density Function}
\newacronym{ppa}{PPA}{Power, Performance, Area}
\newacronym{prng}{PRNG}{Pseudo-Random Number Generator}
\newacronym{ram}{RAM}{Random Access Memory}
\newacronym{rcg}{RCG}{Rocket Chip Generator}
\newacronym{rf}{RF}{Radio Frequency}
\newacronym{rgb}{RGB}{Red/Green/Blue}
\newacronym{rhs}{RHS}{Right Hand Side}
\newacronym{risc}{RISC}{Reduced Instruction Set Computer}
\newacronym{rnn}{RNN}{Recurrent Neural Network}
\newacronym{rtl}{RTL}{Register Transfer Language}
\newacronym{rv}{RISC-V}{RISC-Five}
\newacronym{sa}{SA}{Simulated Annealing}
\newacronym{sbsn}{SBSN}{Stochastic Bit-Stream Neuron}
\newacronym{sgb}{SGB}{Stochastic Gradient Boosting}
\newacronym{simd}{SIMD}{Single Instruction Multiple Data}
\newacronym{soc}{SoC}{System-on-Chip}
\newacronym{sql}{SQL}{Structured Query Language}
\newacronym{sv}{SV}{System Verilog}
\newacronym{svd}{SVD}{Singular Value Decomposition}
\newacronym{svg}{SVG}{Scalable Vector Graphics}
\newacronym{svhn}{SVHN}{Street View House Numbers}
\newacronym{svm}{SVM}{Support Vector Machine}
\newacronym{tl}{TileLink}{TileLink On-Chip Interconnect}
\newacronym{tlb}{TLB}{Translation Lookaside Buffer}
\newacronym{ts}{TS}{Time Series}
\newacronym{ua}{uarch}{Micro Architecture}
\newacronym{uob}{UoB}{University of Bristol}
\newacronym{usb}{USB}{Universal Serial Bus}
\newacronym{ust}{UltraSoC}{UltraSoC Technologies Ltd}
\newacronym{v95}{V95}{Verilog (1995)}
\newacronym{vcd}{VCD}{Value Change Dump}
\newacronym{vhdl}{VHDL}{Very High Speed Integrated Circuit Hardware Description Language}
\newacronym{vliw}{VLIW}{Very Long Instruction Word}
\newacronym{vlsi}{VLSI}{Very Large Scale Integration}
\newacronym{xml}{XML}{eXtensible Markup Language}
\newacronym{yaml}{YAML}{YAML Ain't Markup Language}

\newacronym{TP}{TP}{True-Positive}
\newacronym{TN}{TN}{True-Negative}
\newacronym{FP}{FP}{False-Positive}
\newacronym{FN}{FN}{False-Negative}
\newacronym{TPR}{TPR}{True Positive Rate (Sensitivity)} 
\newacronym{TNR}{TNR}{True Negative Rate (Specificity)} 
\newacronym{PPV}{PPV}{Positive Predictive Value (Precision)}
\newacronym{NPV}{NPV}{Negative Predictive Value}
\newacronym{FNR}{FNR}{False Negative Rate}
\newacronym{FPR}{FPR}{False Positive Rate}
\newacronym{FDR}{FDR}{False Discovery Rate}
\newacronym{FOR}{FOR}{False Omission Rate}
\newacronym{ACC}{ACC}{Accuracy}
\newacronym{BACC}{BACC}{Balanced Accuracy}
\newacronym{MCC}{MCC}{Matthews Correlation Coefficient}
\newacronym{BMI}{BMI}{Book-Maker's Informedness}

\usepackage{subcaption} 

\usepackage{breakurl}

\usepackage{amsmath,amssymb,amsfonts}
\usepackage{mathtools}
\usepackage{relsize} 
\DeclareMathOperator{\NaN}{NaN}
\newcommand{\indep}{\perp\!\!\!\perp} 
\DeclareMathOperator{\Ex}{\mathbb E} 
\DeclareMathOperator{\cov}{cov} 

\renewcommand{\leq}{\leqslant} 

\DeclareMathOperator{\sEx}{\mathbb E} 
\DeclareMathOperator{\sHam}{\dot{H}am}

\DeclareMathOperator{\sTmt}{\dot{T}mt}
\DeclareMathOperator{\sCls}{\dot{C}ls}
\DeclareMathOperator{\sCos}{\dot{C}os}
\DeclareMathOperator{\sDep}{\dot{D}ep}
\DeclareMathOperator{\sCov}{\dot{C}ov}

\usepackage{pifont}

\usepackage[binary-units]{siunitx}

\DeclareSIUnit{\nothing}{\relax} 

\let\DeclareUSUnit\DeclareSIUnit

\DeclareUSUnit\ounce{oz} 
\DeclareUSUnit\inch{in} 
\DeclareUSUnit\foot{ft} 
\DeclareUSUnit\mil{mil} 

\DeclareSIUnit\dBi{dBi} 

\usepackage{listings}

\lstdefinelanguage{none}{
  identifierstyle=
}

\lstdefinelanguage{SystemVerilog}%
  {morekeywords={
      always_comb,always_ff,always_latch,%
      bit,logic,int,var,%
      always,and,assign,automatic,begin,buf,bufif0,bufif1,case,casex,%
      casez,cell,cmos,config,deassign,default,defparam,design,disable,%
      edge,else,end,endcase,endconfig,endfunction,endgenerate,%
      endmodule,endprimitive,endspecify,endtable,endtask,event,for,%
      force,forever,fork,function,generate,genvar,highz0,highz1,if,%
      ifnone,incdir,include,initial,inout,input,instance,integer,join,%
      large,liblist,library,localparam,macromodule,medium,module,nand,%
      negedge,nmos,nor,noshowcancelled,not,notif0,notif1,or,output,%
      parameter,pmos,posedge,primitive,pull0,pull1,pulldown,pullup,%
      pulsestyle_onevent,pulsestyle_ondetect,rcmos,real,realtime,reg,%
      release,repeat,rnmos,rpmos,rtran,rtranif0,rtranif1,scalared,%
      showcancelled,signed,small,specify,specparam,strong0,strong1,%
      supply0,supply1,table,task,time,tran,tranif0,tranif1,tri,tri0,%
      tri1,triand,trior,trireg,unsigned,use,vectored,wait,wand,weak0,%
      weak1,while,wire,wor,xnor,xor},%
   morekeywords=[2]{
      $clog2,$onehot,%
      $bitstoreal,$countdrivers,$display,$fclose,$fdisplay,$fmonitor,%
      $fopen,$fstrobe,$fwrite,$finish,$getpattern,$history,$incsave,%
      $input,$itor,$key,$list,$log,$monitor,$monitoroff,$monitoron,%
      $nokey},%
   morekeywords=[3]{
      `accelerate,`autoexpand_vectornets,`celldefine,`default_nettype,%
      `define,`else,`elsif,`endcelldefine,`endif,`endprotect,%
      `endprotected,`expand_vectornets,`ifdef,`ifndef,`include,%
      `no_accelerate,`noexpand_vectornets,`noremove_gatenames,%
      `nounconnected_drive,`protect,`protected,`remove_gatenames,%
      `remove_netnames,`resetall,`timescale,`unconnected_drive},%
   alsoletter=\`,%
   sensitive,%
   morecomment=[s]{/*}{*/},%
   morecomment=[l]//,
   morestring=[b]"%
  }[keywords,comments,strings]%

\newcommand*{\fancyreflstlabelprefix}{lst}

\fancyrefaddcaptions{english}{%
  \providecommand*{\freflstname}{listing}%
  \providecommand*{\Freflstname}{Listing}%
}

\frefformat{plain}{\fancyreflstlabelprefix}{\freflstname\fancyrefdefaultspacing#1}
\Frefformat{plain}{\fancyreflstlabelprefix}{\Freflstname\fancyrefdefaultspacing#1}

\frefformat{vario}{\fancyreflstlabelprefix}{%
  \freflstname\fancyrefdefaultspacing#1#3%
}

\Frefformat{vario}{\fancyreflstlabelprefix}{%
  \Freflstname\fancyrefdefaultspacing#1#3%
}

\usepackage{cite} 

\hypersetup{
    colorlinks,
    filecolor=black,
    citecolor=red,
    linkcolor=red,
    urlcolor=blue
}

\begin{document}

\title{\doctitle
  \thanks{This project is supported by the Engineering and Physical Sciences
  Research Council (EP/I028153/ and EP/L016656/1);
  the University of Bristol and UltraSoC Technologies Ltd.
  }
}

\author{
  Dave M\textsuperscript{c}Ewan\orcidID{0000-0002-1125-2022}
  \and Jose Nunez-Yanez\orcidID{0000-0002-5153-5481}
}

\authorrunning{D. M\textsuperscript{c}Ewan et al.}

\institute{
  University of Bristol, UK \email{\{dave.mcewan,eejlny\}@bristol.ac.uk}
}

\maketitle

\begin{abstract}
\gls{soc} designs are used in every aspect of computing and their optimization
is a difficult but essential task in today's competitive market.
Data taken from \gls{soc}s to achieve this is often characterised by very long
concurrent bit vectors which have unknown relationships to each other.
This paper explains and empirically compares the accuracy of several methods
used to detect the existence of these relationships in a wide range of systems.
A probabilistic model is used to construct and test a large number of
\gls{soc}-like systems with known relationships which are compared with the
estimated relationships to give accuracy scores.
The metrics $\sCov$ and $\sDep$ based on covariance and independence are
demonstrated to be the most useful, whereas metrics based on the Hamming
distance and geometric approaches are shown to be less useful for detecting the
presence of relationships between \gls{soc} data.

\end{abstract}


\section{Introduction} 
\label{sec:introduction}

\gls{soc} designs include the processors and associated peripheral blocks of
silicon chip based computers and are an intrinsic piece of modern computing,
owing their often complex design to lifetimes of work by hundreds of hardware
and software engineers.
The \gls{soc} in a RaspberryPi\cite{BCM2836} for example includes 4 ARM
processors, memory caches, graphics processors, timers,
and all of the associated interconnect components.
Measuring, analysing, and understanding the behavior of these systems is
important for the optimization of cost, size, power usage, performance, and
resiliance to faults.

Sampling the voltage levels of many individual wires is typically infeasible due
to bandwidth and storage constraints so sparser event based measurements are
often used instead;
E.g. Observations like ``\texttt{cache\_miss} @ \SI{123}{\nano\second}''.
This gives rise to datasets of very long concurrent streams of binary
occurrence/non-occurrence data so an understanding of how these event
measurements are related is key to the design optimization process.
It is therefore desirable to have an effective estimate of the connectedness
between bit vectors to indicate the existence of pairwise relationships.
Given that a \gls{soc} may perform many different tasks the relationships
may change over time which means that a windowed or, more generally,
a weighted approach is required.
Relationships between bit vectors are modelled as boolean functions composed
of negation (NOT), conjunction (AND), inclusive disjunction (OR),
and exclusive disjunction (XOR) operations since this fits well with natural
language and has previously been successfully applied to many different system
types\cite{MeasureLogicComplexity};
E.g. Relationships of a form like ``\texttt{flush} occurs when
\texttt{filled} AND \texttt{read\_access} occur together''.

This paper provides the following novel contributions:
\begin{itemize}
\item A probabilistic model for \gls{soc} data which allows a large amount of
    representative data to be generated and compared on demand.
\item An empirical study on the accuracy of several weighted correlation and
    similarity metrics in the use of relationship estimation.
\end{itemize}
A collection of previous work is reviewed, and the metrics are formally
defined with the reasoning behind them.
Next, assumptions about the construction of \gls{soc} relationships are
explained and the design of the experiment is described along with the method
of comparison.
Finally results are presented as a series of \gls{pdf} plots and discussed in
terms of their application.


\section{Previous Work} 
\label{sec:previouswork}

An examination of currently available hardware and low-level software
profiling methods is given by Lagraa\cite{LagraaThesis} which covers well known
techniques such as using counters to generate statistics about both
hardware and software events -- effectively a low cost data compression.
Lagraa's thesis is based on profiling \gls{soc}s created specifically on
Xilinx MPSoC devices, which although powerful, ensures it may not be applied to
data from non-\gls{fpga} sources such as designs already manufactured in
silicon which is often the end goal of \gls{soc} design.
Lo et al\cite{MiningPastTemporalRules} described a system for describing
behavior with a series of statements using a search space exploration process
based on boolean set theory.
While this work has a similar goal of finding temporal dependencies it is
acknowledged that the mining method does not perform adequately for the very
long traces often found in real-world \gls{soc} data.
Ivanovic et al\cite{TSAnalysisPossApp} review time series
analysis models and methods where characteristic features of economic time
series are described such as drawn from noisy sources, high auto-dependence
and inter-dependence, high correlation, and non-stationarity.
\gls{soc} data is expected to have these same features, together with
full binarization and much greater length.
The expected utility approach to learning probabilistic models by Friedman and
Sandow\cite{LearningProbabilisticModels} minimises the Kullbach-Leibler
distance between observed data and a model, attempting to fit that data using
an iterative method.
As noted in Friston et al\cite{BayesModelReduct} fully learning all parameters
of a Bayesian network through empirical observations is an intractable analytic
problem which simpler non-iterative measures can only roughly approximate.
The approach of modelling relationships as boolean functions has been used for
measuring complexity and pattern detection in a variety of fields including
complex biological systems from the scale of proteins to groups of
animals\cite{InfoProcLivingSystems}.

`Correlation' is a vague term which has several possible
interpretations\cite{ThirteenWaysCorrelationCoefficient} including treating
data as high dimensional vectors, sets, and population samples.
A wide survey of binary similarity and distance measures by
Choi et al\cite{SurveyBinarySimilarityMeasures}
tabulates 76 methods from various fields and classify them as either distance,
non-correlation, or correlation based.
A similarity measure is one where a higher result is produced for more similar
data, whereas a distance measure will give a higher results for data which are
further apart, i.e, less similar.
The distinction between correlation and similarity can be shown with an example:
If it is noticed over a large number of parties that the pattern of attendance
between Alice and Bob is similar then it may be inferred that there is some kind
of relationship connecting them.
In this case the attendance patterns of Alice and Bob are both similar and
correlated.
However, if Bob is secretly also seeing Eve it would be noticed that Bob only
attends parties if either Alice or Eve attend, but not both at the same time.
In this case Bob's pattern of attendance may not be similar to that of either
Alice or Eve, but will be correlated with both.
It can therefore be seen that correlation is a more powerful approach for
detecting relationships, although typically involves more calculation.

In a \gls{soc} design the functionallity is split into a number of discrete
logical blocks such as a timer or an ARM processor which communicate via one
or more buses.
The configuration of many of these blocks and buses is often specified with a
non-trivial set of parameters which affects the size, performance, and cost of
the final design.
The system components are usually a mixture of hardware and software which
should all be working in harmony to achieve the designer's goal and the designer
will usually have in mind how this harmony should look.
For example the designer will have a rule that they would like to confirm
``software should use the cache efficiently'' which will be done by analysing
the interaction of events such as \texttt{cache\_miss} and
\texttt{enter\_someFunction}.
By recording events and measuring detecting inter-event relationships the system
designer can decide if the set of design parameters should be kept or
changed\cite{paper0}, thus aiding the \gls{soc} design optimization process.


\section{Metrics} 
\label{sec:metrics}

A measured stream of events is written as $f_i$ where $i$ is an identifier for
one particular event source such as \texttt{cache\_miss}.
Where $f_i(t) = 1$ indicates event $i$ was observed at time $t$, and $f_i(t) = 0$
indicating $i$ was not observed at time $t$.
A windowing or weighting function $w$ is used to create a weighted average of
each measurement to give an expectation of an event occurrence.
\begin{align}
\label{eq:def_Ex}
{\sEx}{\left[ f_i \right]} &= \frac{1}{\sum_{t} w(t)}
    \mathlarger{\sum}_{t} w(t) * f_i(t)
\quad \in [0,1]
\end{align}

Bayes theorem may be rearranged to find the conditional expectation.
\begin{align}
\label{eq:bayes} 
\Pr(X|Y) &= \frac{\Pr(Y|X) \Pr(X)}{\Pr(Y)} = \frac{\Pr(Y \cap X)}{\Pr(Y)},
    \quad \text{if}\ \Pr(Y) \neq 0
\\
\label{eq:def_cEx}
{\sEx}{\left[ f_x | f_y \right]} &:=
    \begin{dcases}
    \NaN &: {\sEx}{\left[ f_y\right]} = 0
    \\
    \frac{{\sEx}{\left[ f_x * f_y \right]}}
         {{\sEx}{\left[ f_y \right]}}
        &: \text{otherwise}
    \end{dcases}
\end{align}
It is not sufficient to look only at conditional expectation to determine if $X$
and $Y$ are related.
For example, the result $\Pr(X|Y) = 0.9$ may arise from $X$'s relationship with
$Y$, but may equally arise from the case $\Pr(X) = 0.9$.

A na\"ive approach might be to estimate how similar a pair of bit vectors
are by counting the number of matching bits.
The expectation that a pair of corresponding bits are equal is the Hamming
Similarity\cite{Hamming1950}, as shown in \fref{eq:def_Ham}.
Where $X$ and $Y$ are typical sets\cite{InfoMacKay} this is equivalent to
$\left| {\Ex}{\left[ X \right]} - {\Ex}{\left[ Y \right]} \right|$.
The absolute difference $\left| X - Y \right|$ may also be performed on binary
data using a bitwise XOR operation.
\begin{align}
\label{eq:def_Ham}
\sHam(f_x, f_y) &:= 1 - {\sEx}{\left[ \left| f_x - f_y \right| \right]}
\end{align}
The dot in the notation is used to show that this measure is similar to, but not
necessarily equivalent to the standard definition.
Modifications to the standard definitions may include disallowing $\NaN$,
restricting or expanding the range to $[0,1]$, or reflecting the result.
For example, reflecting the result of
${\sEx}{\left[ \left| f_x - f_y \right| \right]}$ in the definition of $\sHam$
a metric is given where $0$ indicates fully different and $1$ indicates exactly
the same.

A similar approach is to treat a pair of bit vectors as a pair of
sets.
The Jaccard index first described for comparing the distribution of alpine
flora\cite{JaccardAlpineFlora}, and later refined for use in general sets
is defined as the ratio of size the intersection to the size of the union.
Tanimoto's reformulation\cite{TanimotoClassifyPlants} of the Jaccard index
shown in \Fref{eq:tanimoto} was given for measuring the similarity of binary
sets.
\begin{align}
\label{eq:tanimoto}
J(X, Y) &= \frac{\left| X \cap Y \right|}
                {\left| X \cup Y \right|}
         = \frac{\left| X \cap Y \right|}
                {\left| X \right| + \left| Y \right| - \left| X \cap Y \right|}
,\quad \left| X \cup Y \right| \neq \varnothing
\\
\label{eq:def_Tmt}
\sTmt(f_x, f_y) &:=
    \frac{ {\sEx}{\left[ f_x * f_y \right]} }
         { {\sEx}{\left[ f_x \right]} + {\sEx}{\left[ f_y \right]} - {\sEx}{\left[ f_x * f_y \right]} }
\end{align}

Treating measurements as points in bounded high dimensional space allows the
Euclidean distance to be calculated, then reflected and normalized to $[0,1]$
to show closeness rather than distance.
This approach is common for problems where the alignment of physical objects is
to be determined such as facial detection and gene
sequencing\cite{ClusteringByPassingMessages}.
\begin{align}
\label{eq:def_Cls}
\sCls(f_x, f_y) &:= 1 -
    \sqrt{{\sEx}{\left[ \left| f_x - f_y \right|^2 \right]}}
\end{align}
It can be seen that this formulation is similar to using the Hamming distance,
albeit growing quadratically rather than linearly as the number of identical
bits increases.
Another geometric approach is to treat a pair of measurements as bounded high
dimensional vectors and calculate the angle between them using the cosine
similarity as is often used in natural language processing\cite{Word2Vec}
and data mining\cite{SemanticCosineSimilarity}.
\begin{align}
\label{eq:cosinesimilarity}
\text{CosineSimilarity}_{X,Y} &=
    \frac{X \cdot Y}{\left| X \right| \left| Y \right|}, \ X,Y \neq 0
\quad \in [-1,1]
\\
\label{eq:def_Cos}
\sCos(f_x, f_y) &:=
    \frac{ {\sEx}{\left[ f_x * f_y \right]} }
         {\sqrt{ {\sEx}{\left[ f_x^2 \right]} } \sqrt{ {\sEx}{\left[ f_y^2 \right]} }}
\quad \in [0,1]
\end{align}
The strict interval of the measured bit vectors $f_x, f_y \in [0,1]$ mean that
$\sCos$ is always positive.

The above metrics attempt to uncover relationships by finding pairs of
bit vectors which are similar to each other.
These may be useful for simple relationships of forms similar to
``\texttt{X} leads to \texttt{Y}'' but may not be useful for finding
relationships which incorporate multiple measurements via a function of boolean
operations such as
``\texttt{A} AND \texttt{B} XOR \texttt{C} leads to \texttt{Y}''.
Treating measurement data as samples from a population invites the use of
covariance or the Pearson correlation coefficient as a distance metric.
The covariance, as shown in \Fref{eq:covariance}, between two bounded-value
populations is also bounded, as shown in \Fref{eq:covariance_limits}.
This allows the $\sCov$ metric to be defined, again setting negative correlations
to $0$.
For binary measurements with equal weights $\sCov$ can be shown to be
equivalent to the Pearson correlation coefficient.
\begin{align}
\label{eq:covariance}
\cov(X, Y) &= {\Ex}{\left[
    \left(X-{\Ex}{\left[ X \right]}\right)
    \left(Y-{\Ex}{\left[ Y \right]}\right)
                    \right]}
    = {\Ex}{\left[ XY \right]} - {\Ex}{\left[ X \right]}{\Ex}{\left[ Y \right]}
\\
\label{eq:covariance_limits}
X,Y \in [0,1] &\implies \frac{-1}{4} \leq \cov(X,Y) \leq \frac{1}{4}
\\
\label{eq:def_Cov}
\sCov(f_x, f_y) &:=
            4\ \Big|
                {\sEx}{\left[ f_x * f_y \right]} -
                {\sEx}{\left[ f_x \right]}{\sEx}{\left[ f_y \right]}
              \Big|
\quad \in [0,1]
\end{align}
Using this definition it can be seen that if two random variables are
independent then $\sCov(X,Y) = 0$, however the reverse is not true in general
as the covariance of two dependent random variables may be $0$.
The definition of independence in \Fref{eq:independence} may be used to define
a metric of dependence.
\begin{align}
\label{eq:independence}
X \indep Y &\iff \Pr(X) = \Pr(X|Y)
\\
\label{eq:def_Dep}
\sDep(f_x, f_y) &:=
        \Bigg|
        \frac{{\sEx}{\left[ f_x | f_y \right]} - {\sEx}{\left[ f_x \right]}}
                                 {{\sEx}{\left[ f_x | f_y \right]}}
        \Bigg|,
    \quad \text{if}\ {\sEx}{\left[ f_x \right]} \leq {\sEx}{\left[ f_x | f_y \right]}
\end{align}
Normalizing the difference in expectation
${\sEx}{\left[ f_x | f_y \right]} - {\sEx}{\left[ f_x \right]}$
to the range $[0,1]$ allows this to be rearranged showing that $\sDep(X,Y)$
is an undirected similarity, i.e. the order of $X$ and $Y$ is unimportant.
\begin{align}
\sDep(f_x, f_y) &=
    \frac{{\sEx}{\left[ f_x | f_y \right]} - {\sEx}{\left[ f_x \right]}}
         {{\sEx}{\left[ f_x | f_y \right]}}
    = 1 - \frac{{\sEx}{\left[ f_x \right]} {\sEx}{\left[ f_y \right]}}
               {{\sEx}{\left[ f_x * f_y \right]}}
    = \sDep(f_y, f_x)
\end{align}

The metrics defined above $\sHam$, $\sTmt$, $\sCls$, $\sCos$,
$\sCov$, and $\sDep$ all share the same codomain $[0,1]$ where $1$ means
the strongest relationship.
In order to compare these correlation metrics an experiment has been devised
to quantify their effectiveness, as described in \fref{sec:experiment}.


\section{Experimental Procedure} 
\label{sec:experiment}

This experiment constructs a large number of \gls{soc}-like systems according to
a probabilistic structure and records event-like data from them.
The topology of each system is fixed which means relationships between bit
vectors in each system are known in advance of applying any estimation metric.
The metrics above are then applied to the recorded data and compared to the
known relationships which allows the effectiveness of each metric to be
demonstrated empirically.

\begin{figure}[t] 
\centering
\includegraphics[width=0.6\linewidth]{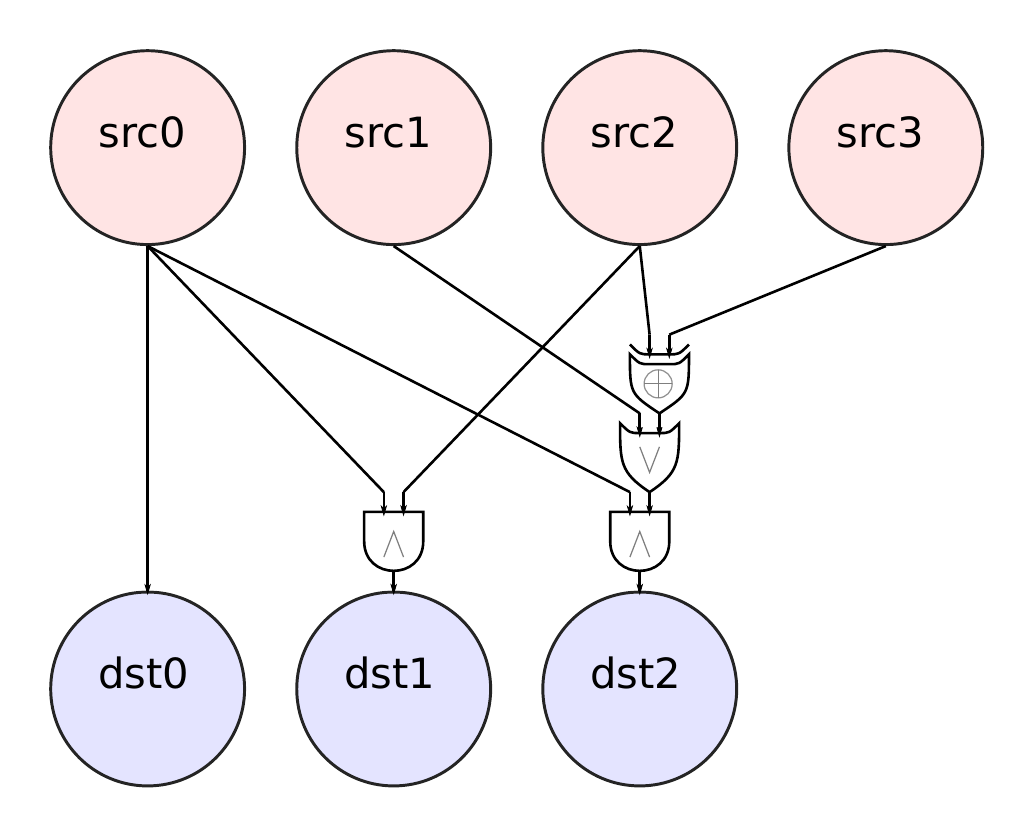}
\caption{Example system with src and dst nodes connected via binary operations.
\label{fig:eg_soc_relest}}
\end{figure}

The maximum number of measurement nodes $2n_{\text{maxm}}$ is set to $100$ to
keep the size of systems within reasonable limits.
Each system is composed of a number of measurement nodes $e_{i \in [1, m]}$ such
that $m = m_{\text{src}} +  m_{\text{dst}}$ of either type `src' or `dst'
arranged in a bipartite graph as shown in \Fref{fig:eg_soc_relest}.
In each system the numbers of measurement nodes are chosen at random
$m_{\text{src}}, m_{\text{dst}} \sim {\operatorname{U}}{(1, n_{\text{maxm}})}$.
Src nodes are binary random variables with a fixed densitity
$\sim {\operatorname{Arcsin}}{(0,1)}$ where the approximately equal number of
high and low density bit vectors represents equal importance of detecting
relationships and anti-relationships.
The value of each dst node is formed by combining a number of edges
$\sim {\operatorname{Lognormal}}{(0,1)}$ from src nodes.
There are five types of systems which relate to the method by which src nodes
are combined to produce the value at a dst node.
One fifth of systems use only AND operations ($\land$) to combine connections
to each dst node, another fifth uses only OR ($\lor$), and another fifth uses
only XOR ($\oplus$).
The fourth type of system uniformly chooses one of the $\land$, $\lor$, $\oplus$
methods to give a mix of homogeneous functions for each dst node.
The fifth type gets the values of each dst node by applying chains of
operations $\sim {\operatorname{U}}{( \{ \land, \lor, \oplus \} )}$ combine
connections, implemented as \gls{lha}.
By keeping different connection strategies separate it is easier to see how the
metrics compare for different types of relationships.

The known relationships were used to construct an adjacency matrix where
$K_{ij} = 1$ indicates that node $i$ is connected to node $j$, with $0$
otherwise.
The diagonal is not used as these tautological relationships will provide a
perfect score with every metric without providing any new information about the
metric's accuracy or effectiveness.
Each metric is applied to every pair of nodes to construct an estimated
adjacency matrix $E$.
Each element $E_{ij}$ is compared with $K_{ij}$ to give an amount of
\gls{TP} and \gls{FN} where $K_{ij} = 1$ or an amount of
\gls{TN} and \gls{FP} where $K_{ij} = 0$.
For example if a connection is known to exist ($K_{ij} = 1$) and the metric
calculated a value of $0.7$ then the \acrlong{TP} and \acrlong{FN} values
would be $0.7$ and $0.3$ respectively, with both \acrlong{TN} and
\acrlong{FP} equal to $0$.
Alternatively if a connection is know to not exist ($K_{ij} = 0$) then
\acrlong{TN} and \acrlong{FP} would be $0.3$ and $0.7$, with \acrlong{TP}
and \acrlong{FN} equal to $0$.
These are used to construct the confusion matrix and subsequently give scores
for the \gls{TPR}, \gls{TNR}, \gls{PPV}, \gls{NPV}, \gls{ACC}, \gls{BACC},
\gls{BMI}, and \gls{MCC}.
\begin{minipage}[t]{0.50\textwidth}
\begin{align*}
\text{TP} &= \sum_{i,j} {\min}{\left( K_{ij}, E_{ij} \right)} \\
\text{FP} &= \sum_{i,j} {\min}{\left( 1-K_{ij}, E_{ij} \right)} \\
\text{TPR} &= \frac{\text{TP}} {\text{TP} + \text{FN}} \\
\text{PPV} &= \frac{\text{TP}} {\text{TP} + \text{FP}} \\
\text{ACC} &= \frac{\text{TP} + \text{TN}}
                   {\text{TP} + \text{FN} + \text{TN} + \text{FP}} 
\end{align*}
\end{minipage}%
\begin{minipage}[t]{0.50\textwidth}
\begin{align*}
\text{FN} &= \sum_{i,j} {\min}{\left( K_{ij}, 1-E_{ij} \right)} \\
\text{TN} &= \sum_{i,j} {\min}{\left( 1-K_{ij}, 1-E_{ij} \right)} \\
\text{TNR} &= \frac{\text{TN}} {\text{TN} + \text{FP}} \\
\text{NPV} &= \frac{\text{TN}} {\text{TN} + \text{FN}} \\
\text{BACC} &= \frac{\text{TPR} + \text{TNR}} {2} \\
\text{BMI} &= \text{TPR} + \text{TNR} - 1 
\end{align*}
\end{minipage}
\begin{center}
  $ \displaystyle
      \begin{aligned}
\text{MCC} &= \frac{\text{TP} \times \text{TN} - \text{TP} \times \text{TN}}
                   {\sqrt{(\text{TP}+\text{FP})(\text{TP}+\text{FN})(\text{TN}+\text{FP})(\text{TN}+\text{FN})}}
    \end{aligned}
  $
\end{center}

To create the dataset $1000$ systems were generated, with $10000$ samples of
each node taken from each system.
This procedure was repeated for each metric for each system and the \gls{pdf}
of each metric's accuracy is plotted using \gls{kde} to see an overview of
how well each performs over a large number of different systems.


\section{Results and Discussion} 
\label{sec:results}

The metrics defined in \fref{sec:metrics} function as binary classifiers
therefore it is reasonable to compare their effectiveness using some of the
statistics common for binary classifiers noted above.
The \gls{TPR} measures the proportion of connections which are correctly
estimated and the \gls{TNR} similarly measures the proportion of
non-connections correctly estimated.
The \gls{PPV} and \gls{NPV} measures the proportion of estimates which are
correctly estimated to equal the known connections and non-connections.
\gls{ACC} measures the likelihood of an estimation matching a known connection
or non-connection.
For imbalanced data sets \gls{ACC} is not necessarily a good way of scoring the
performance of these metrics as it may give an overly optimistic score.
Normalizing \gls{TP} and \gls{TN} by the numbers of samples gives the
\acrlong{BACC}\cite{PREPMt} which may provide a better score for large systems
where the adjacency matrices are sparse.
\acrlong{MCC} finds the covariance between the known and estimated adjacency
matrices which may also be interpreted as a useful score of metric performance.
Youden's J statistic, also known as \acrlong{BMI} similarly attempts to capture
the performance of a binary classifier by combining the sensitivity and
specifitiy to give the probability of an informed decision.

\begin{figure}[h] 
  \centering
  \begin{subfigure}[t]{0.5\textwidth}
  \includegraphics[width=1.0\linewidth]{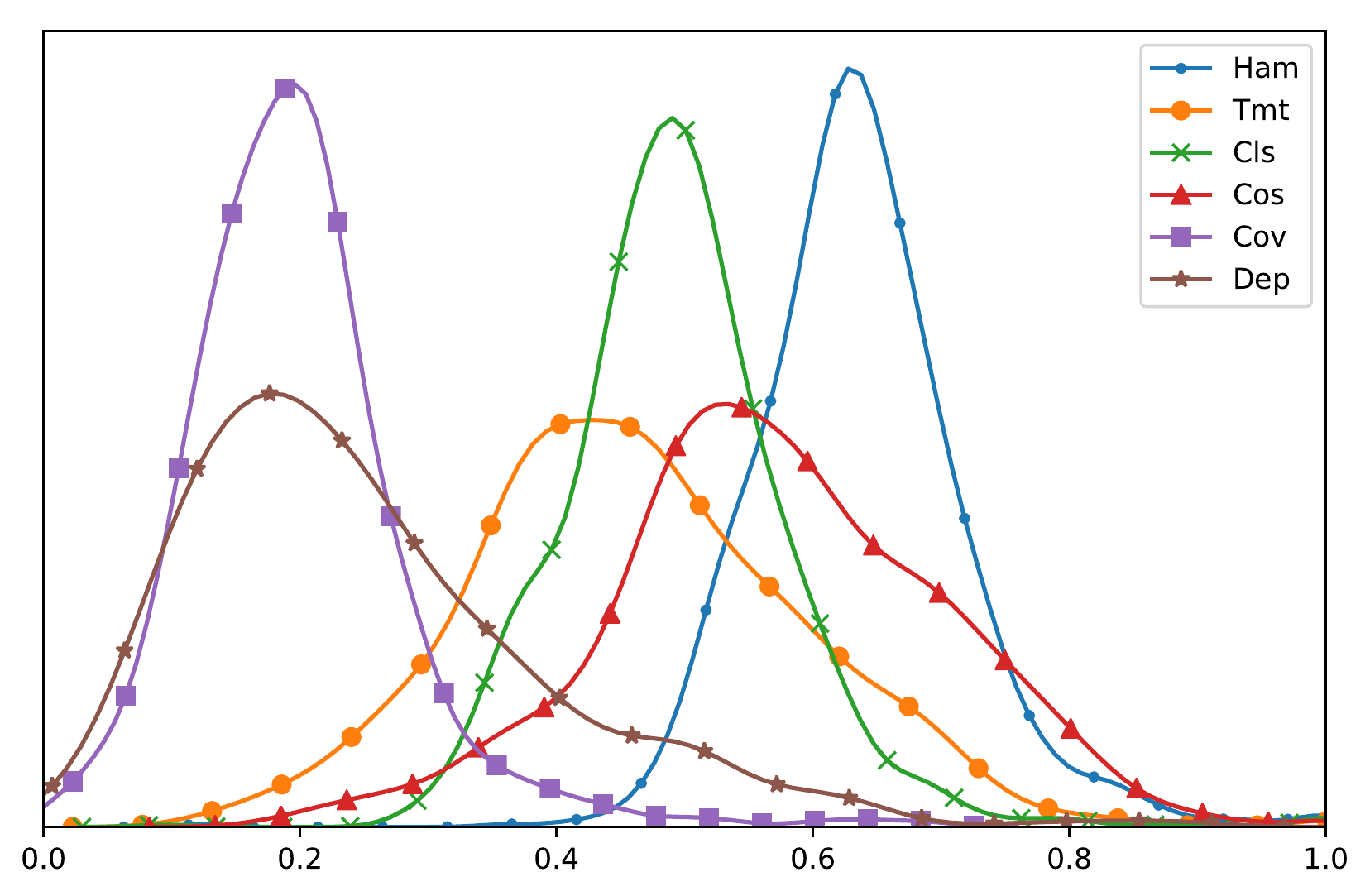}
  \caption{\acrlong{TPR}.
  \label{fig:relest_all_TPR}}
  \end{subfigure}%
  ~
  \begin{subfigure}[t]{0.5\textwidth}
  \includegraphics[width=1.0\linewidth]{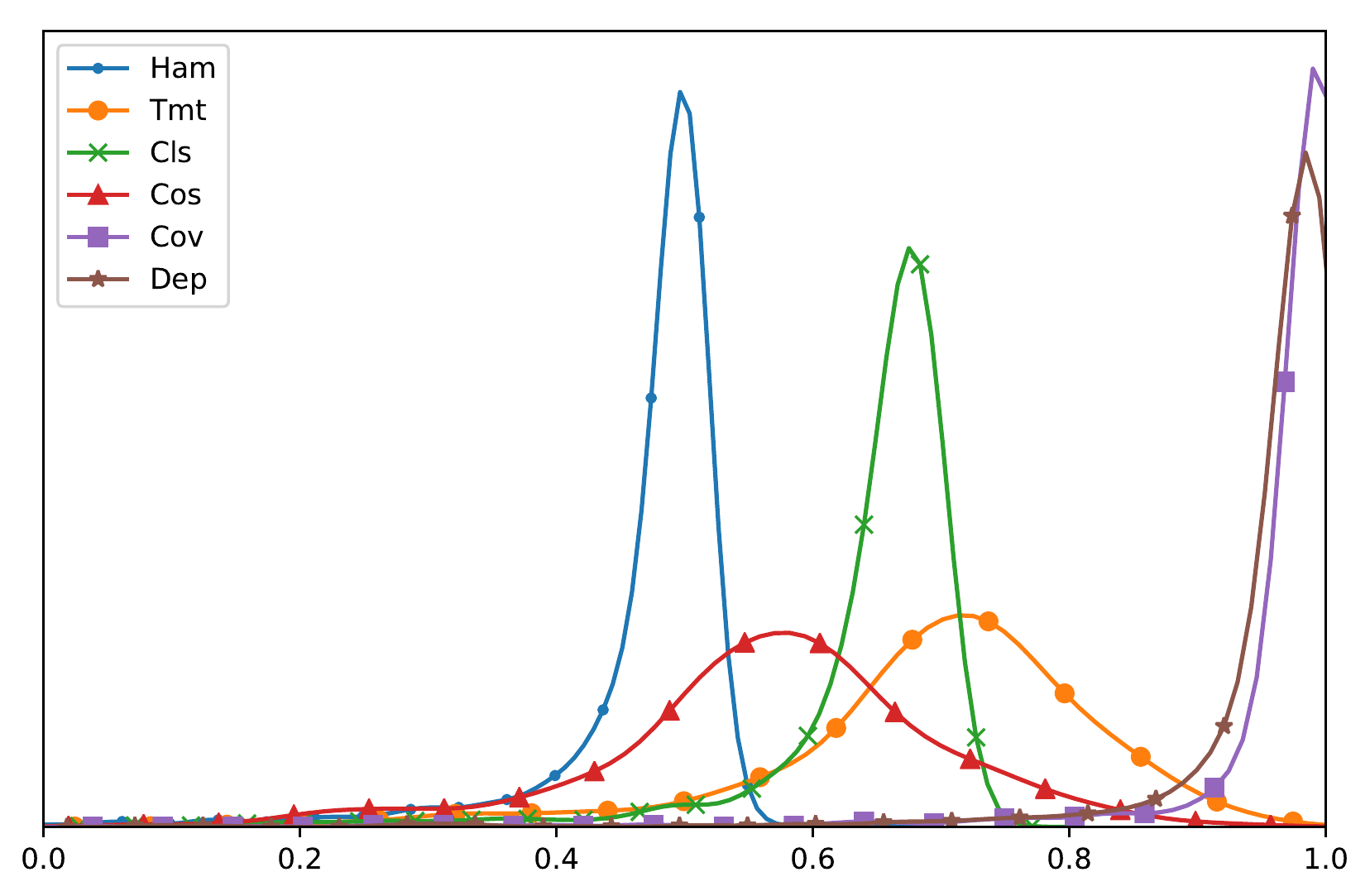}
  \caption{\acrlong{TNR}.
  \label{fig:relest_all_TNR}}
  \end{subfigure}%

  \begin{subfigure}[t]{0.5\textwidth}
  \includegraphics[width=1.0\linewidth]{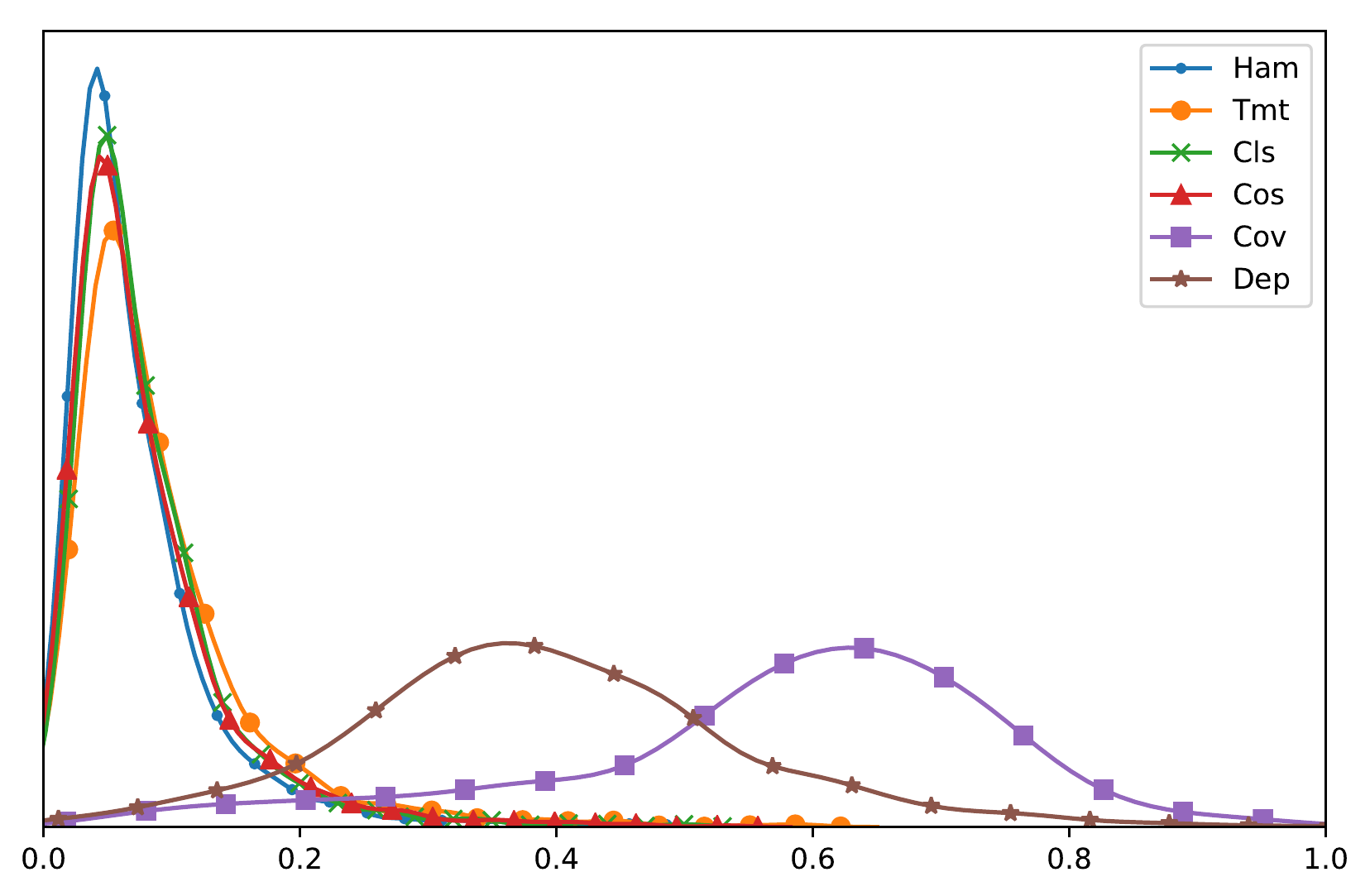}
  \caption{\acrlong{PPV}.
  \label{fig:relest_all_PPV}}
  \end{subfigure}%
  ~
  \begin{subfigure}[t]{0.5\textwidth}
  \includegraphics[width=1.0\linewidth]{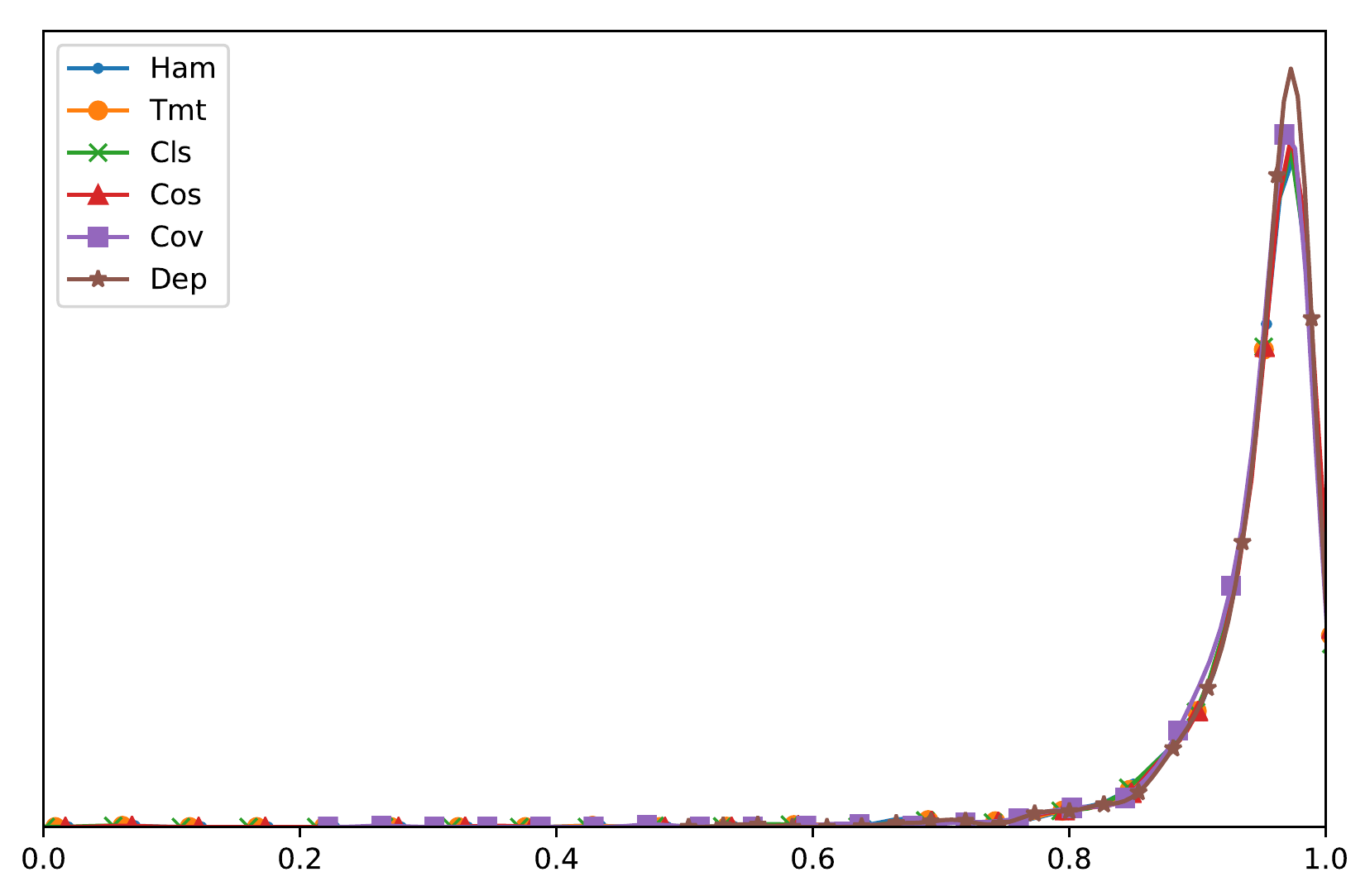}
  \caption{\acrlong{NPV}.
  \label{fig:relest_all_NPV}}
  \end{subfigure}%

  \begin{subfigure}[t]{0.5\textwidth}
  \includegraphics[width=1.0\linewidth]{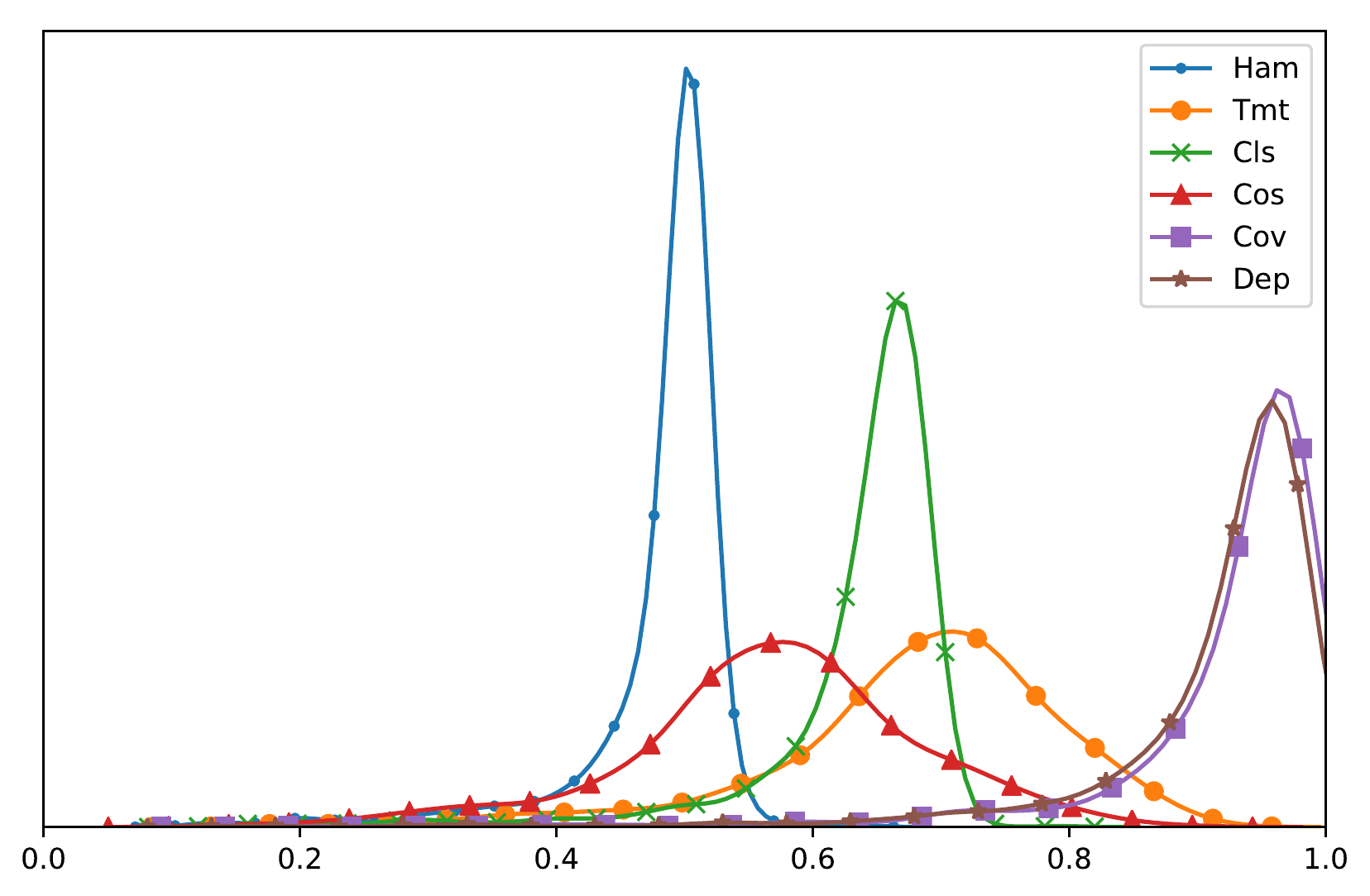}
  \caption{\acrlong{ACC}.
  \label{fig:relest_all_ACC}}
  \end{subfigure}%
  ~
  \begin{subfigure}[t]{0.5\textwidth}
  \includegraphics[width=1.0\linewidth]{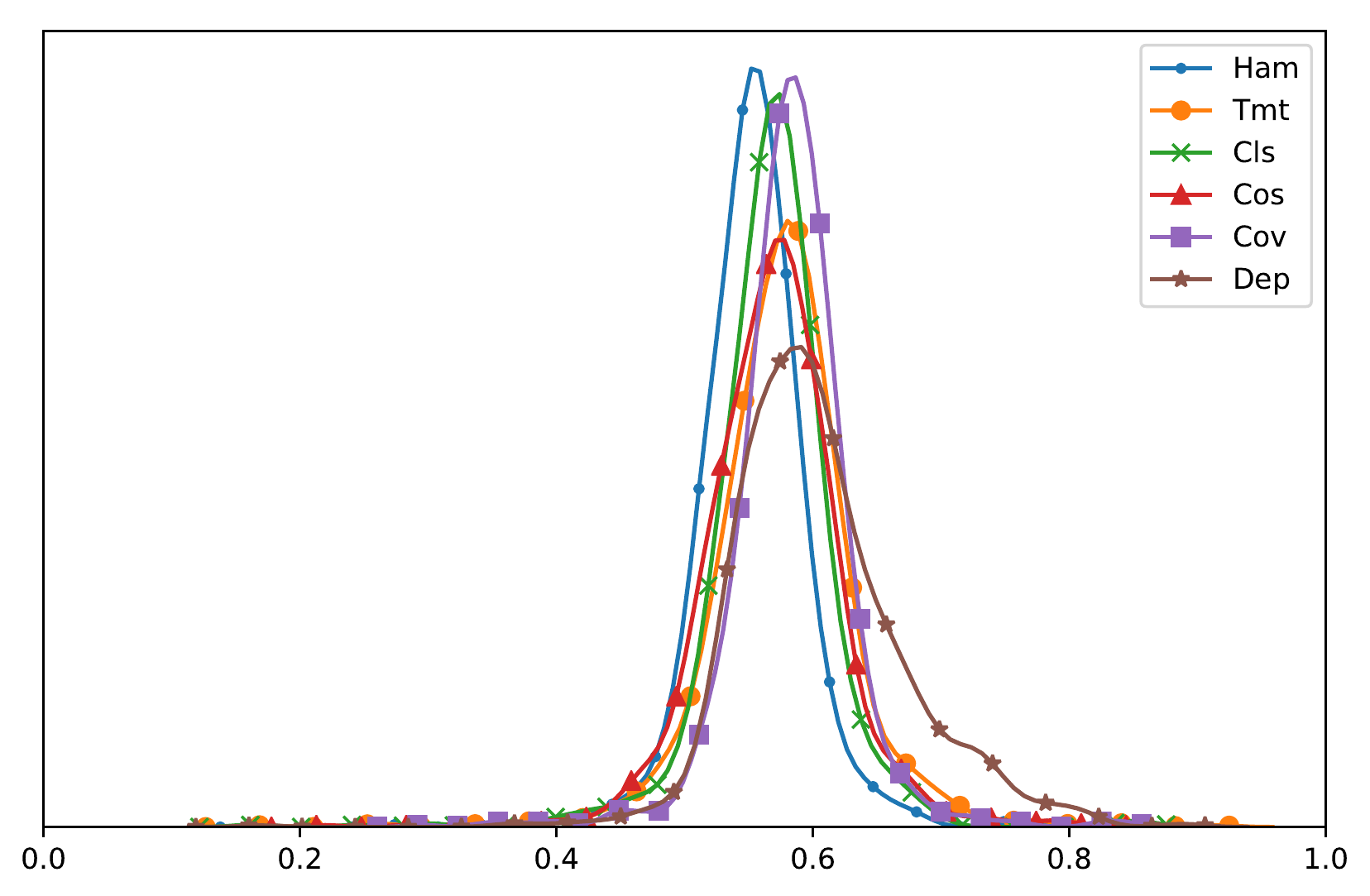}
  \caption{\acrlong{BACC}.
  \label{fig:relest_all_BACC}}
  \end{subfigure}%

  \begin{subfigure}[t]{0.5\textwidth}
  \includegraphics[width=1.0\linewidth]{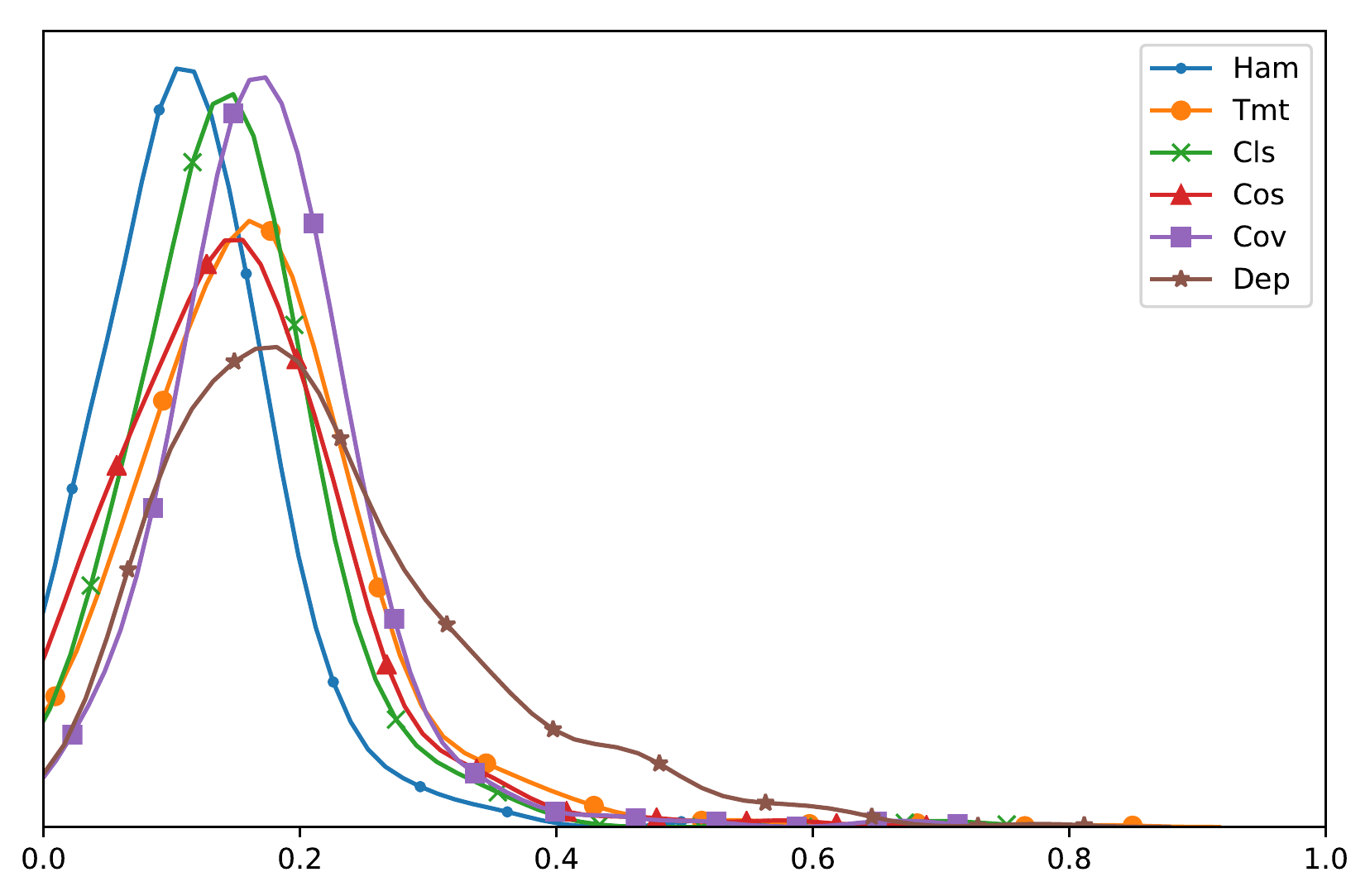}
  \caption{\acrlong{BMI}.
  \label{fig:relest_all_BMI}}
  \end{subfigure}%
  ~
  \begin{subfigure}[t]{0.5\textwidth}
  \includegraphics[width=1.0\linewidth]{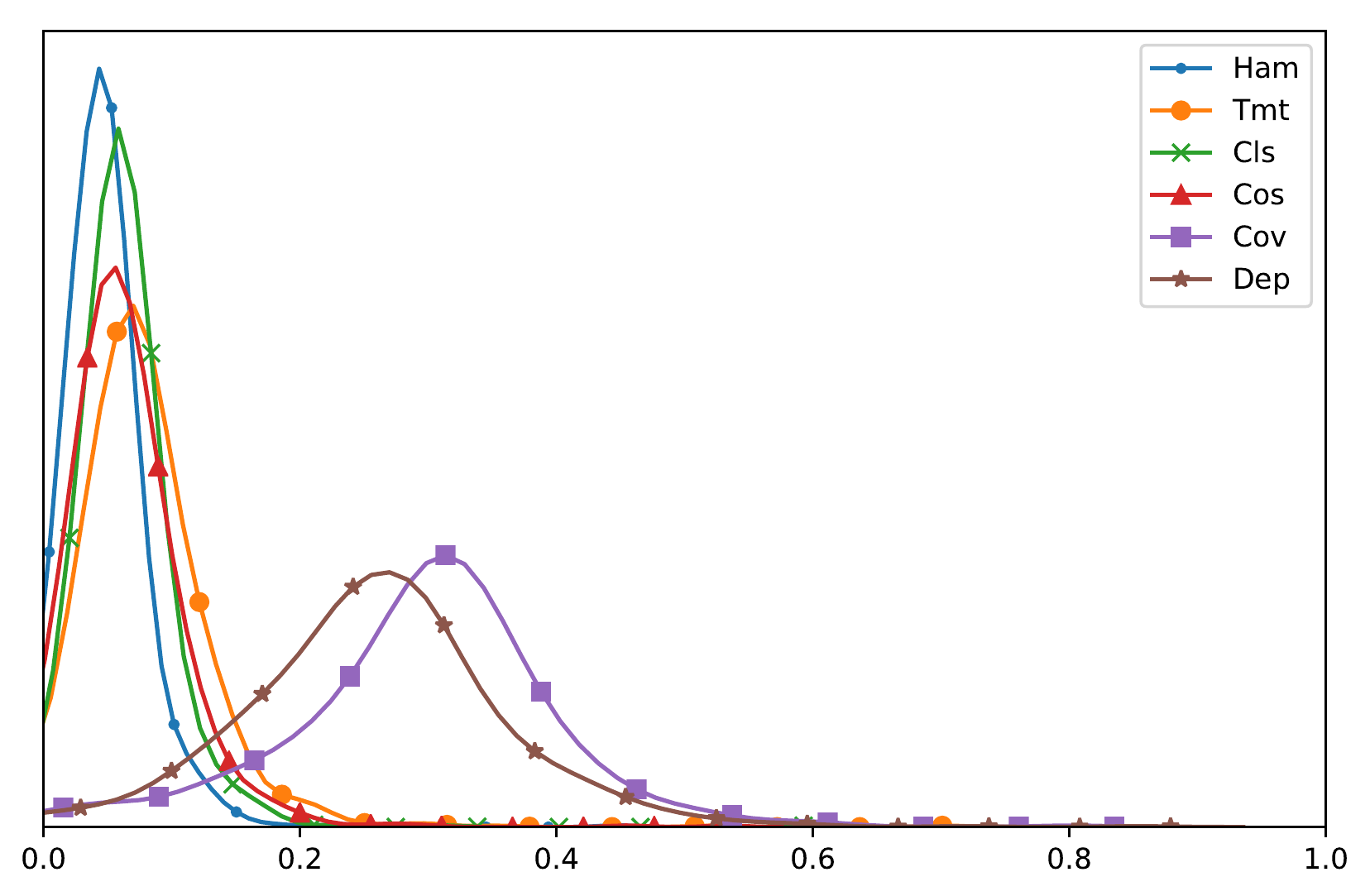}
  \caption{\acrlong{MCC}.
  \label{fig:relest_all_MCC}}
  \end{subfigure}%

  \caption{\gls{kde} plots of score \gls{pdf}s averaged across all system types.
    More weight on the right hand side is always better.
  \label{fig:results1}}
\end{figure}

Each statistic was calculated for each metric for each system.
Given the large number of systems of various types, \glspl{pdf} of these
statistics are shown in \Fref{fig:results1} where more weight on the right
hand side towards $1.0$ indicates a better metric.

\Fref{fig:relest_all_TPR} shows that $\sCov$ and $\sDep$ correctly identify the
existence of around $25\%$ of existing connections and other metrics identify
many more connections.
However, \fref{fig:relest_all_TNR} shows that $\sCov$ and $\sDep$ are much more
likely to correctly identify non-connections than other metrics, especially
$\sHam$ and $\sCov$.

For a metric to be considered useful for detecting connections the expected
value of both \gls{PPV} and \gls{NPV} must be greater than $0$,
and \gls{ACC} must be greater than $0.5$.
It can be seen in \fref{fig:relest_all_NPV} that all metrics score highly for
estimating negatives;
I.e. when a connection does not exist they give a result close to $0$.
On its own this does not carry much meaning as a constant $0$ will always
give a correct answer.
Similarly, a constant $1$ will give a correct answer for positive links
so the plots in the middle and right columns must be considered together with
the overall accuracy to judge the usefulness of a metric.

Given that \gls{ACC} is potentially misleading for imbalanced data sets such as
this one it is essential to check against \gls{BACC}.
$\sHam$ usually has \gls{ACC} of close to $0.5$ which alone indicates than it is
close to useless for detecting connections in binary \gls{soc} data.
The wider peaks of $\sCos$ and $\sTmt$ in both \gls{ACC} and \gls{BACC} indicate
that these metrics are much more variable in their performance than the likes of
$\sCls$, $\sCov$, and $\sDep$.
In this pair of plots where $\sCov$ and $\sDep$ both have much more weight
towards the right hand side this indicates that these metrics are more likely to
give a good estimate of connectedness.

Finally, using \fref{fig:relest_all_MCC} and \fref{fig:relest_all_BMI} as checks
it can be see again that $\sCov$ and $\sDep$ outperform the other metrics.
\gls{MCC} actually has an interval of $[-1,1]$ though the negative side is not
plotted here, and given that all all metrics have weight on the positive side
this shows that all of the defined metrics contain at least some information on
the connectedness.

The overall results indicate that $\sHam$, $\sTmt$, $\sCos$ and $\sCls$ are
close to useless for detecting connections in datasets resembling the \gls{soc}
data model described above.

A characteristic feature employed by both $\sTmt$ and $\sCos$ is the convolution
$f_x * f_y$, whereas $\sHam$ and $\sCls$ employ an absolute difference
$\left| f_x - f_y \right|$.
The best performing metrics $\sCov$ and $\sDep$ have consistently higher
accuracy scores and employ both the convolution, and the product of
expectations ${\sEx}{\left[ f_x \right]} {\sEx}{\left[ f_y \right]}$.

The simplicity of these metrics allows hints about the system function
to be found quickly in an automated manner, albeit without further
information about the formulation or complexity of the relationships.
Any information which can be extracted from a dataset about the workings of its
system may be used to ease the work of a \gls{soc} designer.
For example, putting the results into a suitable visualization provides
an easy to consume presentation of how related a set of measurements are during
a given time window.
This allows the \gls{soc} designer to make a more educated choice about the set
of design parameters in order to provide a more optimal design for their chosen
market.

\section{Conclusion} 
\label{sec:conclusion}

The formulation and rationale behind six methods of measuring similarity or
correlation to estimate relationships between weighted bit vectors has been
given.
The given formulations may also be applied more generally to bounded data in
the range $[0,1]$, though this is not explored in this paper and may be the
subject of future work.
Other directions of future work include testing and comparing additional metrics
or designing specialized metrics for .

It has been shown that using methods which are common in other fields such as
the Hamming distance, Tanimoto distance, Euclidean distance, or Cosine
similarity are not well suited to low-cost relationship detection when the
relationships are potentially complex.
This result highlights a potential pitfall of not considering the system
construction for data scientists working with related binary data streams.

The metrics $\sCov$ and $\sDep$ are shown to consistently estimate the existance
of relationships in \gls{soc}-like data with higher accuracy than the other
metrics.
This result gives confidence that detection systems may employ these approaches
in order to make meaningful gains in the process of optimizing \gls{soc}
behavior.
By using more accurate metrics unknown relationships may be uncovered giving
\gls{soc} designer the information they need to optimize their designs and
sharpen their competitive edge.

The Python code used to perform the experiments is available online\cite{relest}.



\ifdefined\ShowReferences
  \newpage
  \bibliographystyle{splncs04}
  \bibliography{share/refs}{} 
\fi

\ifdefined\ShowGlossary
  \clearpage
  \phantomsection
  \addcontentsline{toc}{section}{Glossary}
  \printnoidxglossary[sort=letter]
\fi


\end{document}